\documentclass[12pt]{article}
\begin{document}
\title{Evidence of a structural anomaly \\
at 14~K in polymerised CsC$_{60}$}
\author{St\'{e}phan Rouzi\`{e}re, Serena Margadonna, \\
Kosmas Prassides and Andrew N. Fitch$^1$ \\
\\
School of Chemistry, Physics and Environmental Science,\\
University of Sussex, Brighton BN1 9QJ, UK\\
$^1$European Synchrotron Radiation Facility, \\
F-38043 Grenoble, France}
\maketitle

\begin{abstract}
We report the results of a high-resolution synchrotron X-ray powder
diffraction study of polymerised CsC$_{60}$ in the temperature range 4 to 40
K. Its crystal structure is monoclinic (space group $I$2/$m$), isostructural
with RbC$_{60}$. Below 14 K, a spontaneous thermal contraction is observed
along both the polymer chain axis, $a$ and the interchain separation along
[111], $d_1$. This structural anomaly could trigger the occurrence of the
spin-singlet ground state, observed by NMR at the same temperature.

PACS.61.48+c Fullerenes and fullerene-related materials.

PACS.61.10Nz Single crystal and powder diffraction.
\end{abstract}

\textit{Introduction.} - Alkali fullerides with stoichiometry AC$_{60}$ ($A$%
= K, Rb, Cs) undergo a structural transition from a high-temperature monomer
to a polymer phase in the vicinity of 350 K \cite{winter,zhu}.
Polymerisation occurs by a [2+2] cycloaddition mechanism, 
leading to the formation of one-dimensional C-C bridged C$_{60}^{-}$
chains \cite{stephens}. The AC$_{60}$ polymers exhibit interesting
structural properties and form a variety of conducting and magnetic
phases. A metal-insulator transition below 50 K is accompanied by the
stabilisation of a magnetic state in RbC$_{60}$ and CsC$_{60}$, whereas KC$%
_{60}$ remains metallic to low temperatures \cite{bommeli}. The nature of
the magnetic transition has remained controversial, as both quasi-one
dimensional electronic instabilities \cite
{bommeli,chauvet,pekker,brouet,janossy} and three-dimensional magnetic
ordering \cite{erwin,auban} have been proposed. Initial X-ray powder
diffraction studies \cite{stephens} have described the structure of the  
polymers as orthorhombic (space group $Pmnn$) with the orientation of
the C$_{60}^{-}$ chains about the short axis $a$ described 
by $\mu $=45$\pm $5${{}^{\circ }}$,
where $\mu $ is the angle between the cycloaddition planes and the $c$ axis.
However, recent single crystal X-ray diffraction and diffuse scattering
studies \cite{launois} on KC$_{60}$ and RbC$_{60}$ have revealed that they
are not isostructural. They adopt different relative chain orientations 
(Figure \ref{fig1}) and their crystal structures are described
by distinct space groups, $Pmnn$ (orthorhombic) and $I$2/$m$ (body-centred
monoclinic), respectively. The difference in their respective physical
properties can thus be attributed to the distinct relative chain
orientations. Moreover the angle $\mu $ was determined as 51${{}^{\circ }}$
for KC$_{60}$ and 47${{}^{\circ }}$ for RbC$_{60}$, the difference also
affecting the electronic band structure \cite{laun2}.

CsC$_{60}$ exhibits overall similar physical properties to RbC$_{60}$.
However, recent NMR measurements have detected the appearance of a
spin-singlet ground state below $T_S$= 13.8 K which coexists with the long
range ordered magnetic state \cite{simovic}. It was proposed that the
development of the non-magnetic phase may be correlated either 
with a structural
change, tentatively ascribed to the occurrence of a spin-Peierls
transition or with an electronic instability.
We note that earlier $\mu$SR data had described the low-temperature state
of CsC$_{60}$ as comprising of co-existing static magnetic and fluctuating
paramagnetic domains \cite{cristofolini}. 
In this paper, we report the crystallographic characterisation
of the CsC$_{60}$ polymer by synchrotron X-ray powder diffraction in the
temperature range 4-40 K. This reveals that CsC$_{60}$ is isostructural with
RbC$_{60}$, adopting a body-centred monoclinic structure (space group $I$2/$m
$). In addition, our results show that a spontaneous strain
appears along both the polymer chain axis and the interchain [111]
direction below $T_S$, providing the signature of magnetoelastic coupling.

\textit{Experimental.} - The CsC$_{60}$ sample was prepared by solid state
reaction of stoichiometric quantities of Cs metal and C$_{60}$ in sealed
quartz tubes at 800 K for four weeks with intermittent shaking.
High-resolution synchrotron X-ray powder diffraction measurements were
performed on a CsC$_{60}$ sample sealed in a 0.5-mm diameter glass
capillary. Data were collected in continuous scanning mode using nine
Ge(111) analyser crystals on the BM16 beamline at the European Synchrotron
Radiation Facility (ESRF), Grenoble, France in the temperature range 4-40 K (%
$\lambda $= 0.79972 \AA ). Data were rebinned in the 2$\theta $ range 5${%
{}^{\circ }}$- 45${{}^{\circ }}$ to a step of 0.01${{}^{\circ }}$. Analysis
of the diffraction data was performed with the GSAS suite of powder
diffraction programmes \cite{GSAS}. In the course of the Rietveld
refinements, the characteristic modulated background was fitted to a
15th-order Chebyshev polynomial. It arises from diffuse scattering, probably
due to the static rotational disorder of residual C$_{60}$ molecules and/or
to the different orientational domains arising from the symmetry lowering at
the monomer- polymer transition. The initial refinements included the scale
factor, the background coefficients, the lattice constants, the zero point
and the peak width parameters. The final refinements incorporated the
isotropic temperature factors, the positional parameters of the bridging C
atoms and the occupation number of Cs.

\textit{Results.} - We first performed detailed Rietveld refinements of the
diffraction profile of CsC$_{60}$ at 20 K. We used three starting structural
models, derived from those reported before for the AC$_{60}$ polymers. In
all cases, the C$_{60}^-$ ions are located at the (0,0,0) and 
($\frac{1}{2}$,$\frac{1}{2}$,$\frac{1}{2}$)
positions, while the Cs$^+$ ions at the (0,0,$\frac{1}{2}$) and 
($\frac{1}{2}$,$\frac{1}{2}$,0) positions
in the unit cell. However, in each case a different polymer chain
orientation ordering is adopted, as illustrated in Figure \ref{fig1}. For
space group $Pmnn$, the orientation of the chains at the origin and the
centre of the unit cell are $\mu$ and -$\mu$, respectively, while for $I$2/$%
m $, they are identical ($\mu$). The third model is described by space group 
$Immm$ and involves a disordered arrangement of chains with $\mu $ or -$\mu $
($\neq$0${{}^{\circ }}$ or 90${{}^{\circ }}$) orientations. In the 
course of the refinements, we monitored the resulting  
quality-of-fit factors ($R_{wp}$) as a function of the orientation of the
chains for all three space groups. Figure \ref{fig2} presents the evolution
of $R_{wp}$ with the rotation angle $\mu$ which was varied between 40${%
{}^{\circ }}$ and 50${{}^{\circ }}$. The refinements were stable throughout
this $\mu$-range with a minimum in all cases at $\mu \simeq $46${{}^{\circ }}
$. However, the deepest minimum is clearly obtained for the monoclinic $I$2/$%
m$ space group, in a similar fashion to RbC$_{60}$ \cite{launois}.
Subsequent refinements concentrated on this structural model and resulted in
excellent quality fits ($R_{wp}$= 3.97\%, $R_{exp}$= 1.44\%). The final results
are plotted in Figure \ref{fig3}. The resulting lattice parameters at 20 K
are $a$= 9.0968(3) \AA, $b$= 10.1895(3) \AA, $c$= 14.1351(4) \AA\ and $%
\alpha $= 89.820(6)${{}^{\circ }}$. The intra- and inter-molecular C$_1$-C$_1
$ bond distances also refined to 1.72(1) \AA\ and 1.60(1) \AA,
respectively, where C$_1$ is the bridging C atom between two C$_{60}$
molecules.

Following the successful determination of the structure of the CsC$_{60}$
polymer at 20 K, we attempted Rietveld refinements of the datasets at other
temperatures. In the course of these refinements, the positional and thermal
atomic parameters and the peak width functions were kept fixed. The
resulting variation of the monoclinic lattice constants in the temperature
range 4-40 K is strongly anisotropic (Figure \ref{fig4}). While the lattice
constants, $b$ and $c$ vary smoothly with temperature, a clear anomaly
occurs below $\approx $14 K along the polymer chain axis $a$. A spontaneous
contraction of the $a$ constant, $\Delta a/a\approx $ 1.5$\times $10$^{-4}$
is evident between 14 and 4 K. At the same time, the monoclinic angle $%
\alpha $ also increases by roughly the same proportion, leading to an
overall contraction in the volume of the unit cell of $\Delta V/V\approx $ 2$%
\times $10$^{-4}$. The observed anomaly is also mirrored in the
interfullerene chain separations. In space group $I$2/$m$, there are two
types of intermolecular environments, $d_{1}$ along the [111] and $d_{2}$
along the [1$\overline{1}$1] directions of the unit cell. These are not
affected equally below 14 K (Figure \ref{fig5}) with a sharp decrease in
interchain separation occurring only for $d_{1}$ along [111] ($\Delta
d_{1}/d_{1}\approx $ 1.0$\times $10$^{-4}$).

\textit{Discussion.} - The temperature-dependent synchrotron X-ray data
have shown that the low-temperature structure of CsC$_{60}$
is identical to that of RbC$_{60}$ with similar orientational angles $\mu$. 
Thus both polymers have comparable
electronic structures \cite{erwin,laun2} and show the appearance of a
low-temperature magnetic insulating phase, in contrast to the non-isostructural
KC$_{60}$ whose metallic state is robust and shows no low-temperature
instabilities. The difference in size between K$^+$, Rb$^+$ and Cs$^+$
also leads to increased nearest-neighbour separations between 
the polymer chains (succesively by $\approx$
0.11\% and 0.12\%) and increased
quasi-one-dimensional character of the electronic structure, as we progress
from KC$_{60}$ to RbC$_{60}$ to CsC$_{60}$. 
In the related family of fulleride polymers, 
Na$_2$Rb$_{1-x}$Cs$_x$C$_{60}$ (0$\leq x \leq$ 1), the electronic
properties were also shown to be very sensitive to such subtle
size effects \cite{arcon}.

The details of the electronic structure of the RbC$_{60}$ and CsC$_{60}$
polymers remain controversial. While many experiments reveal one-dimensional
characteristics \cite{brouet,janossy}, electronic band structure calculations 
predict three-dimensional energy bands \cite{erwin, tanaka}.
In addition, recent $^{13}$C magic angle spinning NMR
experiments on CsC$_{60}$ have shown that the conduction electron density
is concentrated along the equator of the C$_{60}^{-}$ ions and away from
the C-C bridging bonds \cite{swiet}. This indicates that the band structure
is dominated by strong transverse coupling between the polymer chains
rather than a strong one-dimensional coupling along them. Nonetheless,
$^{13}$C and $^{133}$Cs NMR measurements also revealed a transition to
a non-magnetic (spin-singlet) ground state at $T_S$= 13.8 K, ascribed to
a structural or electronic instability \cite{simovic}. The former was 
interpreted as a spin-Peierls transition, consistent with the presence of 
strong 1D features in the electronic description of the CsC$_{60}$ polymer.
 
A highly significant result of the present diffraction experiments is the
observation of a spontaneous thermal contraction along both the chain
axis, $a$ and the interchain separation along [111], $d_1$ below a
transition temperature of about 14 K, which coincides with the spin-singlet
state transition temperature, $T_S$ and unambiguously points to the
structural origin of the instability. This result is certainly 
reminiscent of the
situation encountered for the linear chain compound, CuGeO$_3$ in which the
magnetic transition to a spin-Peierls state is accompanied by shifts of the
Cu$^{2+}$ ions along the chain direction and oxygen displacements perpendicular
to the chains \cite{hirota}; the magnetoelastic coupling was evident in
diffraction experiments
as a spontaneous thermal contraction along the axis $b$, perpendicular
to the chain direction, of comparable magnitude to that observed for
CsC$_{60}$. However, no evidence of any superstructure
peaks is established in CsC$_{60}$, while the quality of the powder diffraction
data is not such as to allow us to determine whether the spontaneous strain
is accompanied by a structural phase transition. 
In this respect, it is also interesting 
to note that the $b$ and $c$ axes parameters of CsC$_{60}$ and the 
interchain distance
along [1$\overline{1}$1], $d_2$ show no anomalies at $T_S$.

The current structural evidence is certainly consistent with the occurrence
of a structural transition in CsC$_{60}$. However, while a spin-Peierls 
scenario is possible, it is equally likely that the observed structural
changes may arise from a CDW (commensurate or incommensurate) transition.
An appealing simple 
interpretation of the structural anomaly at 14 K could be that a soft phonon
mode along the polymer chain may be responsible for the lattice contraction
along $a$ which then induces a local structural rearrangement through 
the variation of the interchain distance $d_1$. In this respect, 
it would be of interest
to perform high-resolution structural measurements at elevated pressures.
The intermolecular distance $d_{1}$ should vary the most, while contraction
along $a$ should be limited because of the rigidity of the polymer
chain. Band structure calculations should then give valuable insight on the
influence of the $d_1$ variation on the nature of the non-magnetic
phase, especially as it has been found experimentally that application of
pressure gradually supresses the magnetic order in favour of the
spin-singlet state in both CsC$_{60}$ and RbC$_{60}$ \cite{simovic,auban}.

\textit{Conclusions.} - In conclusion, high-resolution synchrotron X-ray
powder diffraction has established that the structure of CsC$_{60}$
is monoclinic (space group $I$2/$m$), isostructural with that of RbC$_{60}$. 
The similarity in the orientational ordering of the polymer chains
in these systems explains their comparable electronic and conducting
properties. A structural anomaly is observed below $\approx $14 K, coincident
with the appearance of a non-magnetic state. The observed structural
changes provide the signature of magnetoelastic coupling between the
$C_{60}^{-}$ localised spins and the phonon degrees-of-freedom.
 
{\center ***} \newline
We thank the European Union (TMR Research Network `FULPROP',
ERBFMRXCT970155) and the NEDO Frontier Carbon Technology Program for
financial support and the ESRF for provision of synchrotron X-ray beam time.

\newpage
\begin{figure}
\caption{Schematic drawing of the C$_{60}^-$ chain orientations for space groups
(a) $Pmnn$ and (b) $I$2/$m$. The shaded bars indicate the orientation of
the polymer chains by projection onto the crystallographic
$bc$ plane. The disordered $Immm$ space group discussed in the text corresponds to a random distribution of the chain orientation $\mu/-\mu$
($\neq$0${{}^{\circ }}$ or 90${{}^{\circ }}$).}
\label{fig1}
\end{figure}

\begin{figure}
\caption{Dependence of the reliability factor, $R_{wp}$ of the Rietveld
refinement of the synchrotron X-ray diffraction data of CsC$_{60}$ at 20 K
on the chain orientation $\mu$ for the three space groups $Pmnn$,
disordered $Immm$ and $I$2/$m$.}
\label{fig2}
\end{figure}

\begin{figure}
\caption{Observed (points), calculated (solid line) and difference 
(lower panel) synchrotron X-ray powder diffraction profiles
of CsC$_{60}$ at 20 K (space group $I$2/$m$).
The positions of the reflections are shown as tick marks.}
\label{fig3}
\end{figure}

\begin{figure}
\caption{Temperature dependence of the monoclinic lattice parameters, $a$, $b$, $c$ and $\alpha$ of polymerised CsC$_{60}$. The dotted lines are guides to the eye.}
\label{fig4}
\end{figure}

\begin{figure}
\caption{Temperature dependence of the interchain distances, $d_1$ (along
[111]) and $d_2$ (along [1$\overline{1}$1]) in polymerised CsC$_{60}$. 
The dotted line is a guide to the eye.}
\label{fig5}
\end{figure}


\begin{thebibliography}{99}
\bibitem{winter}  WINTER\ \textit{et al.}, Solid State Commun., \textbf{84}
(1992) 935.
\bibitem{zhu}  ZHU Q. \textit{et al.}, Phys. Rev. B, \textbf{47} (1993)
13948.
\bibitem{stephens}  STEPHENS P. W. \textit{et al.}, Nature, \textbf{370}
(1994) 636.
\bibitem{bommeli}  BOMMELI F. \textit{et al.}, Phys. Rev. B, \textbf{51}
(1995) 14794.
\bibitem{chauvet}  CHAUVET O. \textit{et al.}, Phys. Rev. Lett., \textbf{72}
(1994) 2721.
\bibitem{pekker}  PEKKER S. \textit{et al.}, Solid State Commun., \textbf{90}
(1994) 349.
\bibitem{brouet}  BROUET V. \textit{et al.}, Phys. Rev. Lett., \textbf{76}
(1996) 3638.
\bibitem{janossy}  JANOSSY A. et al., Phys. Rev. Lett., \textbf{79} (1997)
2718; BENNATI M. et al., Phys. Rev. B, \textbf{58} (1998) 15603.
\bibitem{erwin}  ERWIN S. C. \textit{et al.}, Phys. Rev. B, \textbf{51}
(1995) 7345.
\bibitem{auban}  AUBAN-SENZIER P.\textit{\ et al.}, J. Phys. I France, 
\textbf{6} (1996) 2181.
\bibitem{launois}  LAUNOIS P. \textit{et al.}, Phys. Rev. Lett., \textbf{81}
(1998) 4420.
\bibitem{laun2}  LAUNOIS P. \textit{et al.}, Synth. Metals, \textbf{103}
(1999) 2354.
\bibitem{simovic}  SIMOVIC B. \textit{et al.}, Phys. Rev. Lett., \textbf{82}
(1999) 2298.
\bibitem{cristofolini} CRISTOFOLINI L. \textit{et al.}, J. Phys.: Condens. 
Matter, \textbf{7} (1995) L567.
\bibitem{GSAS}  LARSEN A. C. and von DREELE R. B., GSAS software, Los Alamos
National Laboratory Report No. LAUR 86-748.
\bibitem{arcon} ARCON D. \textit{et al.}, Phys. Rev. Lett., \textbf{84}
(2000) 562.
\bibitem{tanaka}  TANAKA K. \textit{et al.}, Chem. Phys. Lett., \textbf{272}
(1997) 189.
\bibitem{swiet}  De SWIET T.M. \textit{et al.}, Phys. Rev. Lett., \textbf{84}
(2000) 717.
\bibitem{hirota} HIROTA K. \textit{et al.}, Phys. Rev. Lett., \textbf{73}
(1994) 736; HARRIS Q. J. \textit{et al.}, Phys. Rev. B, \textbf{50} (1994)
12606.
\end{thebibliography}
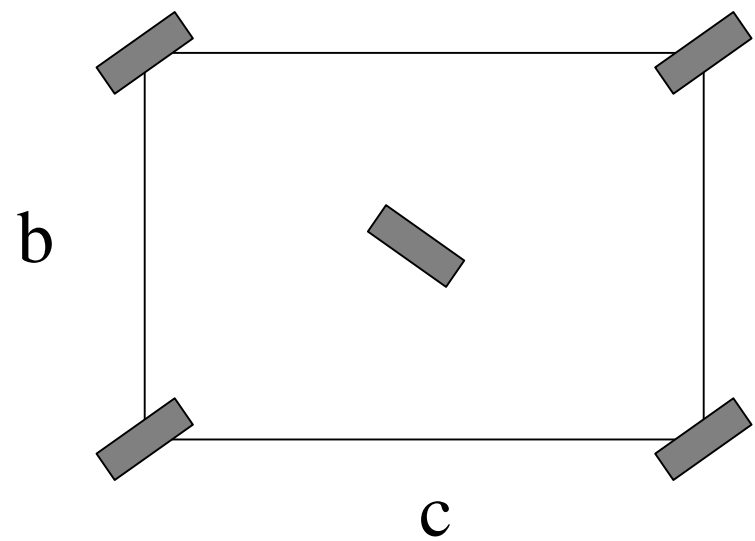
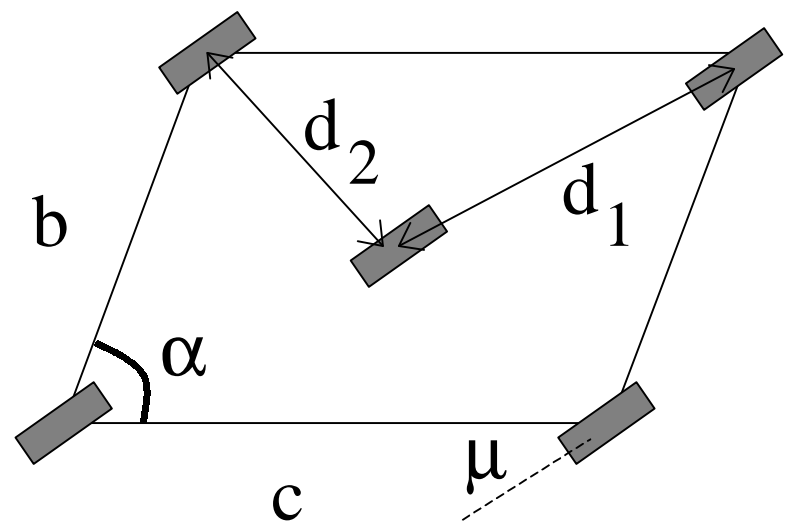
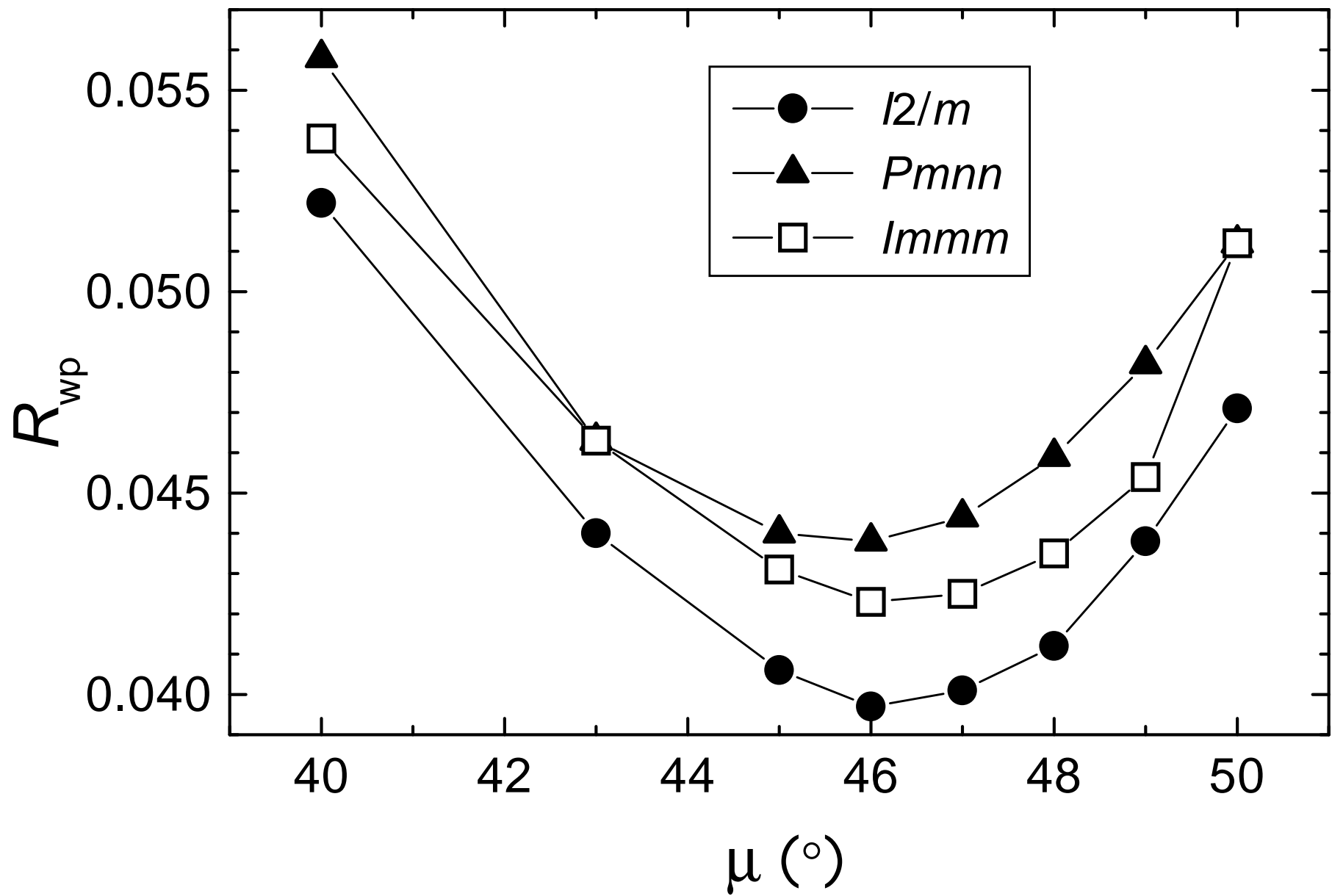
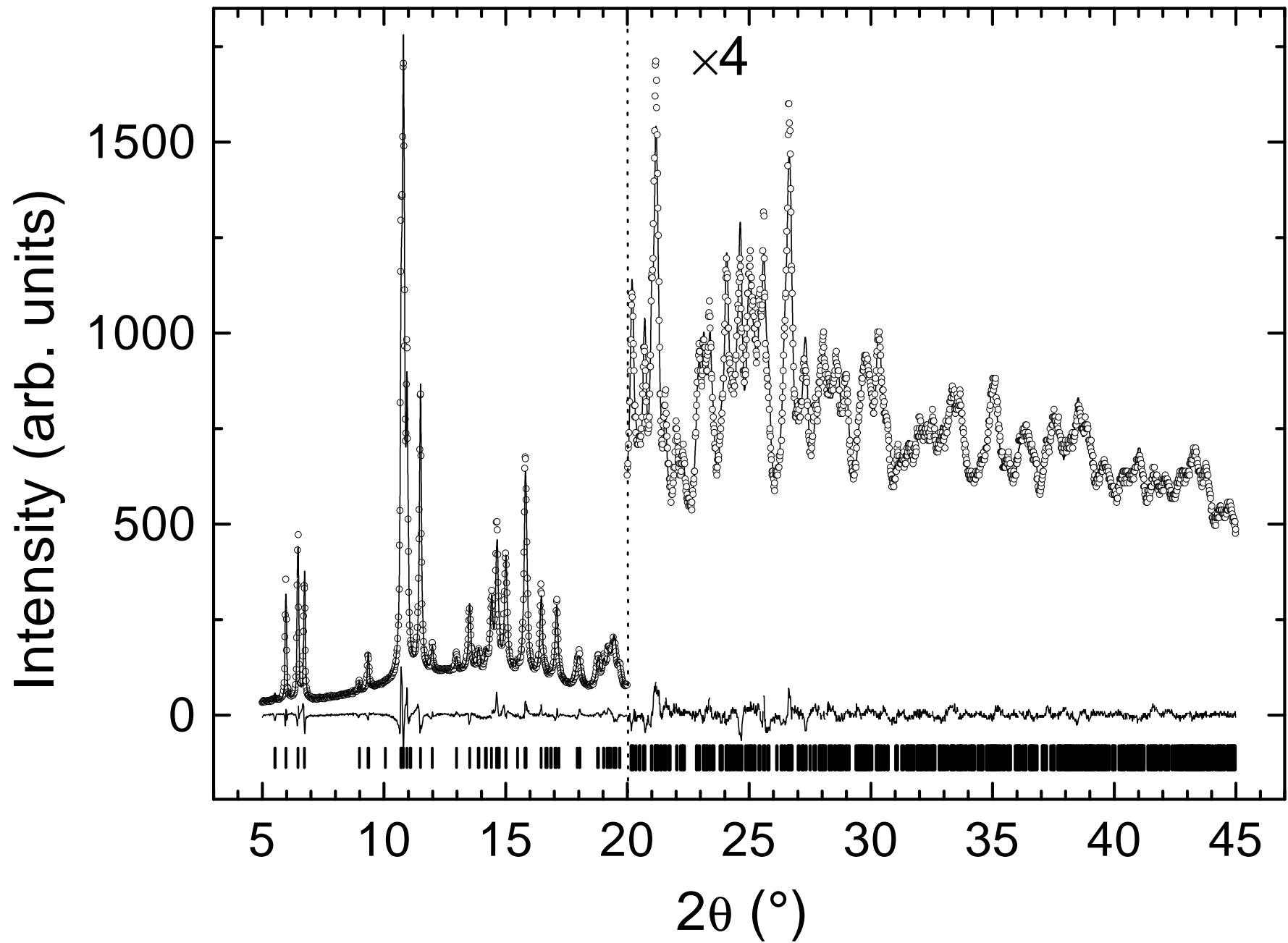
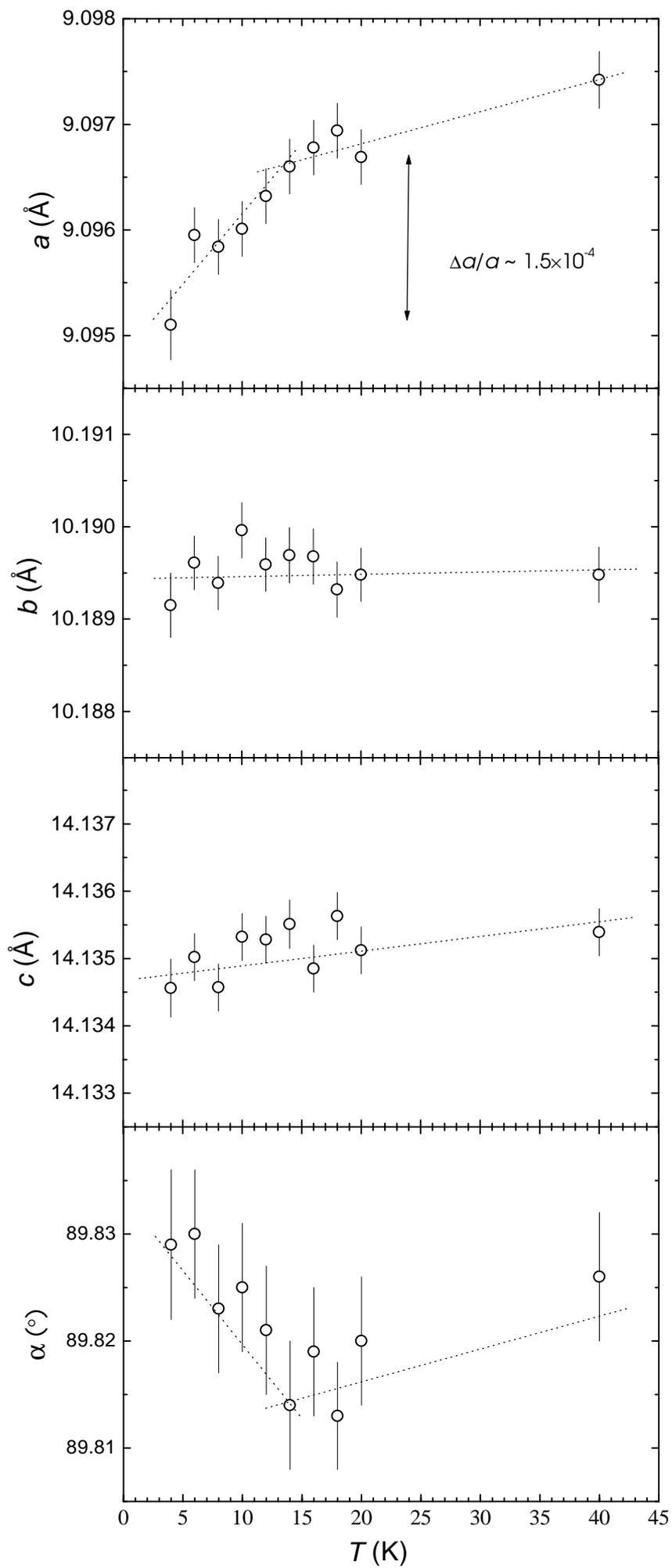
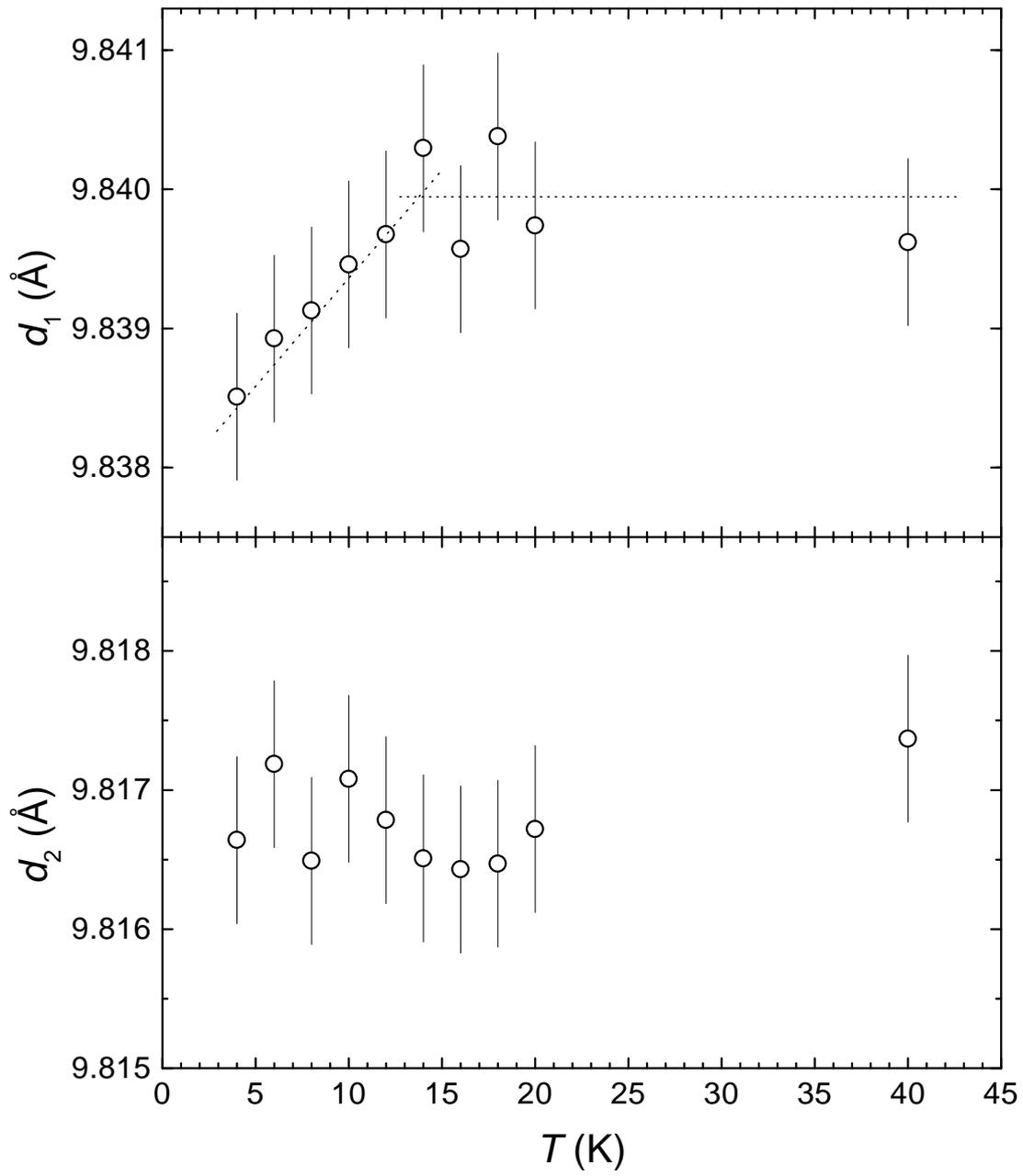
\end{document}